\documentclass[runningheads]{llncs}

\usepackage[margin=1.2in]{geometry}

\usepackage[T1]{fontenc}
\usepackage{comment}
\usepackage{graphicx}

\usepackage[colorlinks,urlcolor=blue]{hyperref}  
\usepackage{color}
\usepackage{url}

\usepackage{caption}

\usepackage{algorithm,algpseudocode}

\usepackage{mathtools}
\usepackage{amsmath,amsfonts,amssymb}
\usepackage{tabularray}
\usepackage{booktabs}
\usepackage{upgreek}
\usepackage{enumitem}
\usepackage{bm}
\usepackage{censor}
\usepackage{textcomp}
\usepackage[english]{babel}
\usepackage{lastpage}
\usepackage[section]{placeins}
\usepackage{fancyvrb} 
\usepackage[dvipsnames,table]{xcolor}
\usepackage{fontawesome5}

\usepackage{multirow}
\usepackage{makecell}
\usepackage{subfigure}

\definecolor{LightCyan}{rgb}{0.88,1,1}
\definecolor{lightskyblue}{RGB}{225, 235, 240}

\definecolor{Gray}{gray}{0.90}
\definecolor{white}{rgb}{1.0, 1.0, 1.0}

\definecolor{Lightgreen}{RGB}{218, 246, 230 }

\usepackage{graphics,color}
\usepackage{xcolor}
\definecolor{label1}{rgb}{0.76,0.59,0.77}
\definecolor{label2}{rgb}{0.28,0.5,0.72}
\definecolor{label3}{rgb}{0.33,0.63,0.36}
\definecolor{label4}{rgb}{0.79,0.4,0.17}
\definecolor{label5}{rgb}{0.94,0.53,0.2}
\definecolor{label6}{rgb}{0.72,0.86,0.59}
\definecolor{label7}{rgb}{1,1,0.65}
\definecolor{label8}{rgb}{0.93,0.62,0.61}
\definecolor{label9}{rgb}{0.4,0.15,0.33}
\definecolor{label10}{rgb}{0.75,0.21,0.29}
\definecolor{label11}{rgb}{0.35,0.73,0.8}
\definecolor{label12}{rgb}{0.94,0.9,0.32}
\definecolor{label13}{rgb}{0.96,0.76,0.48}

\newsavebox{\spleen}
\savebox{\spleen}{\textcolor{label1}{\rule{1.5in}{1.5in}}}

\newsavebox{\rkid}
\savebox{\rkid}{\textcolor{label2}{\rule{1.5in}{1.5in}}}

\newsavebox{\lkid}
\savebox{\lkid}{\textcolor{label3}{\rule{1.5in}{1.5in}}}

\newsavebox{\gall}
\savebox{\gall}{\textcolor{label4}{\rule{1.5in}{1.5in}}}

\newsavebox{\eso}
\savebox{\eso}{\textcolor{label5}{\rule{1.5in}{1.5in}}}

\newsavebox{\liver}
\savebox{\liver}{\textcolor{label6}{\rule{1.5in}{1.5in}}}

\newsavebox{\sto}
\savebox{\sto}{\textcolor{label7}{\rule{1.5in}{1.5in}}}

\newsavebox{\aorta}
\savebox{\aorta}{\textcolor{label8}{\rule{1.5in}{1.5in}}}

\newsavebox{\ivc}
\savebox{\ivc}{\textcolor{label9}{\rule{1.5in}{1.5in}}}

\newsavebox{\veins}
\savebox{\veins}{\textcolor{label10}{\rule{1.5in}{1.5in}}}

\newsavebox{\panc}
\savebox{\panc}{\textcolor{label11}{\rule{1.5in}{1.5in}}}

\newsavebox{\rad}
\savebox{\rad}{\textcolor{label12}{\rule{1.5in}{1.5in}}}

\newsavebox{\lad}
\savebox{\lad}{\textcolor{label13}{\rule{1.5in}{1.5in}}}

\begin{document}

\title{BOrg: A Brain Organoid-Based Mitosis Dataset for Automatic Analysis of Brain Diseases}
\titlerunning{}

\author{Muhammad Awais\textsuperscript{1}, Mehaboobathunnisa Sahul Hameed\textsuperscript{2}, Bidisha Bhattacharya\textsuperscript{3}, Orly Reiner\textsuperscript{4}, Rao Muhammad Anwer\textsuperscript{5}}
\institute{\textsuperscript{1,2,5}Mohamed Bin Zayed University of Artificial Intelligence,\\\textsuperscript{3,4}Weizmann Institute of Science \\
\href{https://awaisrauf.github.io/}{awaisrauf.github.io}, \email{nisa.hameed@mbzuai.ac.ae}, \\ \email{bidisha.bhattacharya@weizmann.ac.il}, \\\email{orly.reiner@weizmann.ac.il}, \email{rao.anwer@mbzuai.ac.ae}}

\maketitle              

\begin{abstract}
Recent advances have enabled the study of human brain development using brain organoids derived from stem cells. Quantifying cellular processes like mitosis in these organoids offers insights into neurodevelopmental disorders, but the manual analysis is time-consuming, and existing datasets lack specific details for brain organoid studies. We introduce BOrg, a dataset designed to study mitotic events in the embryonic development of the brain using confocal microscopy images of brain organoids. BOrg utilizes an efficient annotation pipeline with sparse point annotations and techniques that minimize expert effort, overcoming limitations of standard deep learning approaches on sparse data. We adapt and benchmark state-of-the-art object detection and cell counting models on BOrg for detecting and analyzing mitotic cells across prophase, metaphase, anaphase, and telophase stages. Our results demonstrate these adapted models significantly improve mitosis analysis efficiency and accuracy for brain organoid research compared to existing methods. BOrg facilitates the development of automated tools to quantify statistics like mitosis rates, aiding mechanistic studies of neurodevelopmental processes and disorders. Data and code are available at \href{https://github.com/awaisrauf/borg}{BOrg's GitHub page}.

\keywords{Cell detection, Brain cells, Genetic disorders }
\end{abstract}

\section{Introduction} \label{s:intro}

Understanding early human brain development is crucial for deciphering the mechanisms underlying various neurological conditions like genetic disorders and neurodegenerative diseases. However, directly studying the developing human brain presents significant technical and ethical challenges. Consequently, researchers have heavily relied on animal models, such as rodents, which lack key features of the human brain, including cortical folding believed to be essential for higher cognitive functions~\cite{karzbrun2018human}. This fundamental limitation hinders our ability to fully comprehend human brain development and its associated diseases.

Brain organoids - 3D cell cultures mimicking early brain development - offer a promising alternative to studying the human brain directly \cite{karzbrun2019brain,hofer2021engineering}. This approach avoids ethical concerns and enables researchers to investigate brain development and neurodegenerative diseases in a controlled laboratory setting. However, studying these organoids requires extensive domain expertise and is time-consuming. Manual monitoring of the mitotic process is time-consuming and error-prone, representing a major bottleneck in research. 

\begin{figure}[t]
 \centering
  \includegraphics[width=1\textwidth]
  {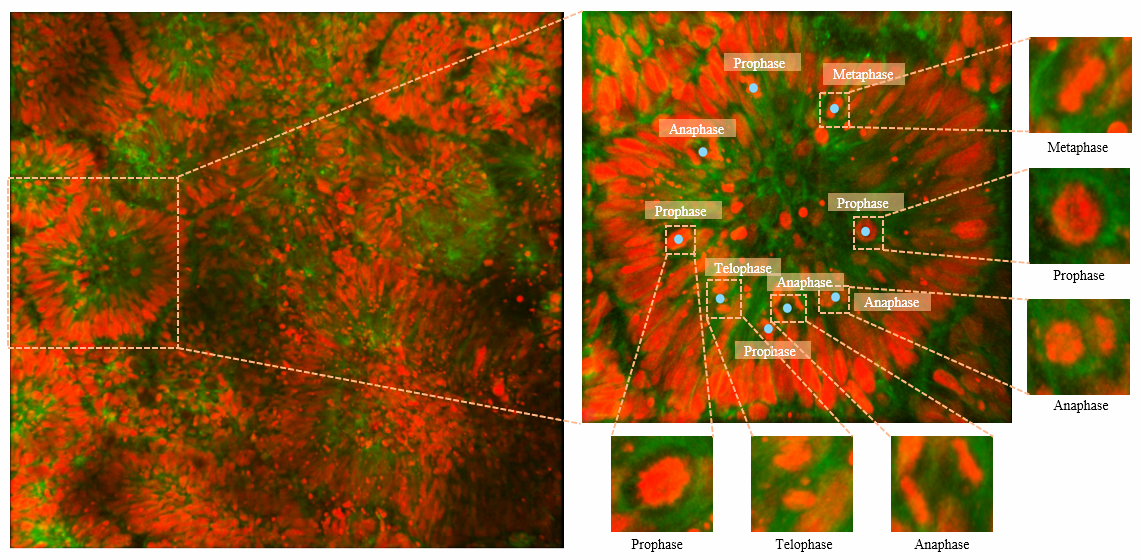} 
  \caption{A sample from the BOrg dataset. \textbf{Left:} a large field-of-view (FoV) patch highlights multiple rosette-like zones in the brain organoids, which mimic the cortical folding observed in the human brain. \textbf{Right:} a detailed close-up of a singular rosette-like germinal zone with point and class annotations as well as a close-up of individual cells in various stages of mitosis.}
  \label{fig:sample_frame}
\vspace{-10pt}
\end{figure}

To address this challenge, we propose a deep learning approach to automate and analyze mitosis in brain organoids, ultimately aiding in understanding brain diseases. Mitosis, the process of cell division, plays a critical role in early brain development. Its proper regulation is crucial; any disruptions can lead to tumor formation or developmental disorders.  
We reformulate the quantification problem into cell identification and counting across different stages of mitosis. The detection and counting information can be used to calculate various statistics and gain deeper insight into brain development and disease.

Our approach involves collecting confocal microscopy images of custom-designed on-chip brain organoids. These brain organoids comprise neural progenitor cells arranged in rosette-like germinal zones that closely mimic early brain development~\cite{karzbrun2018human}. We then developed an efficient data annotation pipeline that utilizes multiple techniques, such as sparse annotations and collaboration between experts and non-experts, to maximize the efficiency of domain expertise. This effort resulted in the creation of the Brain Organoids (BOrg) dataset, consisting of cell images annotated with 737 instances across four mitotic phases: prophase, metaphase, anaphase, and telophase, as depicted in Fig. \ref{fig:sample_frame}. Finally, we adapt state-of-the-art deep learning models for cell detection and counting to detect cells and measure mitosis dynamics. By quantifying these cellular processes, we aim to assist researchers in studying early brain development and deciphering the mechanisms of brain diseases. 

\section{Related Work}
\textbf{Cell Datasets:}
Several public datasets exist for various cell analysis tasks using microscopic images, such as nuclei detection, cell segmentation, detection, and counting. These datasets cover diverse cell types and conditions as presented in \cite{dsb} for the cell segmentation challenge. Additionally, histopathology datasets (e.g., TUPAC16 \cite {tupac}, \cite{consep}, MoNuSAC \cite{monusac}) and related algorithms have been developed for cell segmentation \cite{mitoem}, cell detection \cite{mitnet,spatial_context,gzmh,ocelot},  and cell counting \cite{degpr}.

\noindent\textbf{Cell Detection:} Detecting cells in the biomedical domain is challenging due to their small size and complex appearances. Existing approaches have employed both general purpose and specialized deep learning models such as R-CNN~\cite{rcnn}, Yolo~\cite{yolov3,yolov5,yolov8} and RetinaNet~\cite{retinanet} and DeGPR~\cite{degpr}. Mitosis detection focuses specifically on identifying nuclei undergoing cell division. Several works have explored multi-stage approaches for mitosis detection~\cite{deepmitosis,gzmh}. Similarly, several other works have also been proposed for mitosis detection~\cite{mit_det1,mit_det2,mit_det3,mit_det4}.

\noindent\textbf{Cell Counting:}
General cell counting methods fall under two categories - \textit{detection-based} and \textit{regression-based}. Detection-based methods first localize cells using a detector, followed by a simple counting step. DeGPR~\cite{degpr} introduced a detection-based counting method that detected cells using an improved Yolov5 \cite{degpr} by incorporating supervised contrastive posterior regularization. Similar general object detection models have also been explored \cite{count_detect1,count_detect2}. Regression-based methods formulate cell counting as a regression problem, directly predicting the number of cells~\cite{cell_count_reg1,cell_count_reg2,saunet}. However, this approach has limitations when counting different cell classes.

\section{BOrg Dataset}
The BOrg dataset is a collection of confocal microscopy images of brain organoids. These organoids are specifically designed to mimic the development of the human embryonic brain. The images are systematically captured at regular intervals throughout the development of these organoids with a confocal microscope. Subsequently, each image is efficiently annotated with point annotations corresponding to the mitosis phase, leveraging expert and non-expert skills with an efficient annotation pipeline. Finally, we trained models and demonstrated their effectiveness in detecting and counting cells in different phases. The following section explains our data collection and annotation process in detail in the following section. 

\begin{figure}[t]
    \centering
    \includegraphics[width=0.6\textwidth]{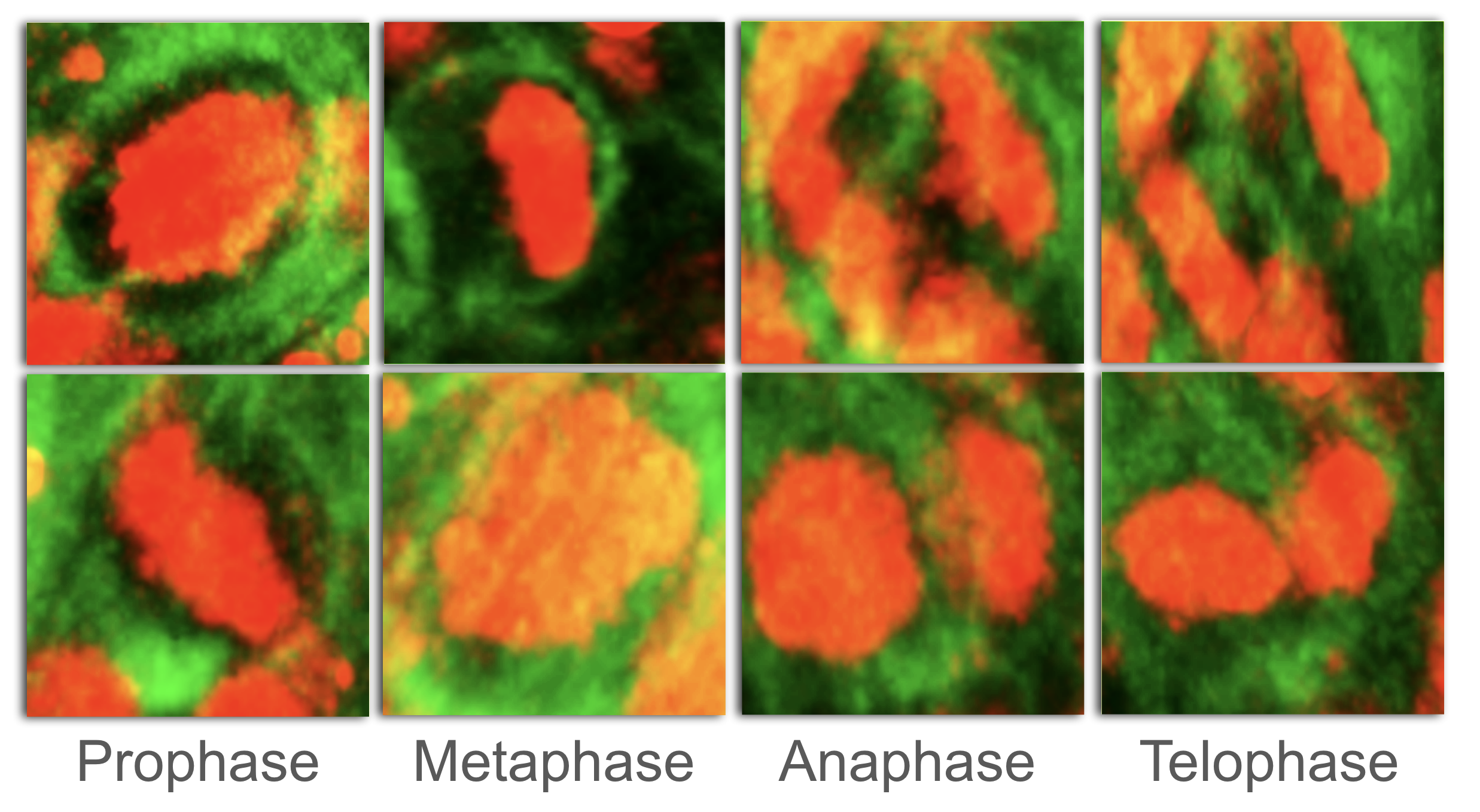}
    \captionof{figure}{Example of cells in four different phases of mitosis. The nuclei are visible with red fluorescence, and the rest of the cell is green fluorescent. }
    \label{fig:mitosis}
\end{figure}




\subsection{Data Collection}
The dataset consists of microscopy images captured at different phases of mitosis in on-chip brain organoids. These brain organoids are composed of neural progenitor cells arranged in rosette-like germinal zones that closely mimic the neuroepithelium stage of early brain development. These neural progenitor cells are proliferating cells that undergo repeated cycles of mitosis to maintain the progenitor cell population and give rise to newly-born neuronal cells. 

These cells have been genetically manipulated such that the nuclei are tagged with red fluorescence, and the rest of the cell is green fluorescent. This helps us to observe the cells during mitosis in real-time by live imaging with confocal microscopy. Fig.~\ref{fig:sample_frame} shows the multiple rosettes in a large FoV and the mitosis events occurring inside each rosette. 
The cell cycle of these proliferating progenitor cells can be divided into two broad phases, like most other dividing cells – the Interphase and Mitosis phases. 

The elongated cells populating the outer edge of the rosettes are in interphase. When the cell decides to enter the mitosis phase, the nuclei become more rounded than elongated and move towards the center of the rosette, where it undergoes all phases of mitosis in succession, namely, Prophase, Metaphase, Anaphase, and Telophase, giving rise to two daughter cells that move back outwards in the rosette. This characteristic cell cycle coordinated movement of the nuclei in neural progenitor cells is called Interkinetic Nuclear Migration (INM) and contributes to normal human brain development. Hence, studying mitosis at this early developmental stage is crucial as it is the foundation of the birthing process of neurons that will later mature and be functional in the brain.

\begin{table}[]
  \centering
  \begin{tabular}{lccc}
  \toprule
    Phase Name & Train Annotations & Validation Annotations & Total Annotations\\
    \toprule
   Prophase & 282 &82  & 364\\
   Metaphase & 146  & 62 & 208\\
   Anaphase & 69  & 24 & 93\\
   Telophase & 59  & 13 & 72\\
    \hline
  \end{tabular}
  \captionof{table}{Annotation statistics and their split. }
  \label{tab:phase_count}
  \vspace{-20pt}
\end{table}

The on-chip brain organoids are generated according to the published protocol from human embryonic stem cell line WIBR3~\cite{karzbrun2018human}. Some genetic manipulations are done to tag the nuclei with red fluorescence and the remaining cells with green fluorescence for real-time tracing of dividing cells. The organoids were allowed to grow till 14 days of development, following which the tissue dynamics were recorded in a confocal microscope. We recorded cell activities for 16 hours at 5-minute intervals using a 40X objective lens in an Oxford Dragonfly spinning disc confocal microscope. The 16-hour movie is split into 16 movies of 1 hour, each containing approximately 20 to 25 frames. Each video shows multiple cells in different phases of mitotic division. These images are then processed and annotated in the files with the Spots feature of Imaris v.6.2. The actual annotations contain the centroid of the cells that undergo mitosis along with their phase of mitotic division.

\subsection{Efficient Data Annotation Pipeline}
Traditionally, analyzing these cellular activities requires extensive expertise in annotating microscopic images. Our work addresses this by introducing an efficient annotation pipeline. This pipeline simplifies the annotation task by focusing on relevant information for the final analysis. It leverages a combination of expert and non-expert under a sparse annotation labeling approach, minimizing the need for manual inputs from domain specialists. Finally, a pre-processing step ensures the data is compatible with training machine learning models. 

In the first stage, we aim to transform the ultimate objective into a task that is efficient in annotating and enables leveraging progress in deep learning by utilizing off-the-shelf models. Given our goal to quantify various statistics, we label each cell undergoing mitosis, intentionally excluding non-dividing cells from this process. Moreover, since our focus is not on precisely detecting cell boundaries, we adopt a strategy of marking dividing cells with dot annotations rather than employing segmentation masks or detection boxes. This approach of sparsely labeling only the cells in the division process substantially conserves expert time.

We implemented a two-stage labeling strategy to expedite the annotation process and alleviate the burden on domain experts. Initially, a domain expert annotates only the start of division for each cell phase. This task necessitates their specialized knowledge due to the significant resemblance between dividing and non-dividing cells and the difficulty for non-experts to distinguish between various cell phases. In the subsequent stage, following the initial expert annotation, a non-expert annotator fills in the remaining frames, thus completing the annotation process. This strategy significantly reduces the required effort; for example, if a cell completes its division over 30 frames, a domain expert needs to annotate only 4 of these.

Upon completing the annotation phase, the final step involves pre-processing the data to ready it for model training. We decided to use 2D projection of 3D data as it enables efficient training and utilizing considerable progress in 2D models. We use mean projection to convert 3D images into 2D images. Finally, by utilizing this framework, we annotated 262 frames taken for 16 hours of microscopy imaging captured every 5 minutes. This way, we collected 737 annotations for four different phases of mitosis. Detailed statistics of our data are shown in Tab.~\ref{tab:phase_count}, and a few examples of cells in different phases of mitosis are shown in Fig.~\ref{fig:mitosis}.

\subsection{Benchmarking by Adapting Object Detection and Cell Counting  Algorithms}
We selected two main categories of models and adapted them for our dataset. These two categories include state-of-the-art (SOTA) off-the-shelf object detection models and algorithms specialized in cell counting. We introduce several modifications to adapt these models for our dataset. These adaptations include converting point annotation to bounding boxes, calculating implicit features, joint pre-training of backbones, etc. 

\begin{table}[t]
    \centering
    \resizebox{\textwidth}{!}{%
    \begin{tabular}{lccccccccccc}
    \toprule
        
        \textbf{Model} &  \textbf{Precision} & \textbf{Recall} & \textbf{mAP} &   \textbf{MAE Pro} &  \textbf{MAE Meta} &  \textbf{MAE Ana} & \textbf{MAE Telo} \\

      \toprule
      FasterRCNN~\cite{faster_rcnn} & 0.364& 0.385 & 0.384 &1.564&1.512&0.615&0.333 \\
     \midrule
      RetinaNet 5~\cite{retinanet}  &0.321 & 0.215 &0.310 & 2.1032 & 1.59 &0.615 &0.333 \\
          \midrule
      YOLOv8~\cite{yolov8}  &0.518 & 0.453 & 0.461 & 1.308 &0.718 & 0.436 &0.359  \\
      YOLOv5~\cite{yolov5}   &0.563 & 0.373 & 0.426 & 1.589 &1.051 &0.436 &0.410 \\
      YOLOv3~\cite{yolov3}   &0.677 & 0.245 & 0.399 &1.103&1.051&0.564 & 0.308  \\ 
            \midrule

     DeGPR++~\cite{degpr} &0.535 & 0.445  &0.449 & 1.487 &0.795 &0.385 &0.308 \\

      \bottomrule
    \end{tabular}
    
    }
    \caption{Comparison of detection and counting performance of various state-of-the-art models adapted for the BOrg dataset. Precision, Recall, mAP (Mean Average Precision), and class-wise MAE (Mean Average Error) are utilized for comparison. DeGPR++ is our adaptation of DeGPR (cell counting algorithm)~\cite{degpr}. }
    \label{tab:detection_results}
    \vspace{-20pt}
\end{table}

\noindent\textbf{Adapting Off-the-Shelf Detection Models for BOrg. }
For our dataset, object detection models can be utilized to compute useful insights. Hence, we first benchmark the BOrg dataset with off-the-shelf object detection algorithms.
As discussed earlier, we transformed dot annotation into detection boxes by utilizing class-wise average cell diameter.
Object detection models require detection bounding boxes rather than the singular point annotations present in our dataset. We computed an average diameter for each cell per class to address this discrepancy and used this value to generate synthetic object bounding boxes. This approach is justified by our objective, which requires the identification rather than the precise detection of different cell types.

Given an image \textit{I} as input, the objective is to predict bounding boxes, $B=\{B_1,B_2,..,B_n\}$ and corresponding classes $c=\{c_1, c_2, ..., c_n\}$. The detector is trained on the ground truth data, where point annotations are converted into box annotations by using the point as the center and drawing a box square box around with an average class diameter. 

To see the effectiveness of different families of object detection methods, we choose three different types: single-stage object detectors~\cite{retinanet}, two-stage object detectors~\cite{faster_rcnn} and YOLO~\cite{yolov3,yolov5,yolov8} family which allows efficient processing on input images. 

\noindent\textbf{Adapting DeGPR for Cell Counting.} In addition to object detection models, we adapt DeGPR~\cite{degpr}, a state-of-the-art cell detection and counting algorithm, to the BOrg dataset. DeGPR formulates cell counting as a detection problem and employs posterior regularization during the training of YOLOv5~\cite{yolov5}. The key idea is to leverage both explicit and implicit discriminative features to train the detector with an additional posterior regularization loss.
Handcrafted attributes such as cell size and intensity are extracted from the bounding box regions for explicit features. A ResNet18 encoder computes implicit features by embedding the cropped bounding box patches. The encoder is trained with a supervised contrastive loss to generate class-discriminative embeddings. These explicit and implicit features are concatenated and fitted with a Gaussian Mixture Model (GMM). KL divergence is applied between the GMM fits of the true and predicted bounding boxes to enforce consistency.
To adapt DeGPR to the BOrg dataset, we first transform the point annotations to bounding boxes using average cell diameters. We then compute explicit handcrafted features from the training set and incorporate them into the model. Finally, we pre-train the embedding encoder on the BOrg dataset to learn discriminative representations tailored to our task. These modifications enable DeGPR to operate effectively on the sparse annotations and unique challenges posed by the BOrg dataset.

\begin{figure}[t]
    \centering
    \includegraphics[width=1\textwidth]{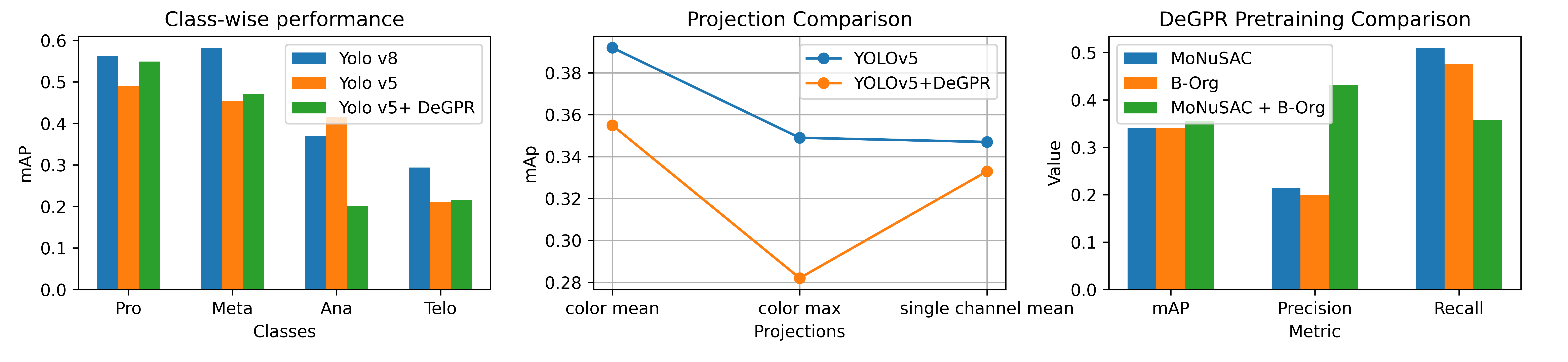}
    \caption{(a) Class-wise mean average precision (mAP) of different detection models. (b) Comparison of the effect of different projections on final mAP. (c) Effect of three different pre-training strategies for DeGPR performance on BOrg dataset. }
    \label{fig:abalations}
\end{figure}
\section{Experiments }

\noindent \textbf{Evaluation metrics. } We utilize detection and counting evaluation metrics since the purpose of our dataset is to calculate various subcellular statistics for later analysis. Specifically, we use precision, recall, and mAP@50 (mean average precision), which measures the accuracy of detection models by considering precision and recall simultaneously, and MAE (mean average error), which quantifies the discrepancy between predicted and ground truth counts of cells at different phases of mitosis. 

\noindent \textbf{Dataset. }
For annotation, we utilize Imaris~\footnote{https://imaris.oxinst.com/}, a microscopy analysis software that enables 3D annotation of cells. In addition to benchmarking our models on our proposed dataset, we also leverage MoNuSAC~\cite{monusac} as a pre-training dataset and demonstrate successful reproduction of DeGPR. MoNuSAC is a histopathology cell dataset comprising four classes (Epithelial, Lymphocyte, Neutrophil, and Macrophage) across 295 images, with 209 for training and 85 for testing. To maintain consistency, we resize each image to 640x640 pixels, irrespective of its original dimensions.
Our novel dataset, BOrg, consists of confocal microscopy frames with an initial size of 654x588 pixels, which we resize to 640x640 for compatibility with our models. The dataset encompasses 262 images, randomly divided into an 80-20 train-validation split (see Tab.~\ref{tab:phase_count}).

\noindent \textbf{Implementation Details.}
For both MoNuSAC and BOrg datasets, we trained detection models for 300 epochs using the SGD optimizer with a training batch size of 32. We employed a learning rate of 0.03 for the MoNuSAC dataset and 0.0005 for the BOrg dataset. In order to find the implicit features for performing posterior regression in DeGPR, the encoder is trained for 300 epochs at a learning rate of 0.001. The model generates 512-dimensional embeddings, which we then reduce using PCA by preserving 90\% of variance.  Additionally, to validate the correct implementation of DeGPR, we initially replicated it on the MoNuSAC dataset, achieving a mean average precision (mAP) close to the reported values (0.481 compared to 0.489). The experiments are conducted on a Quadro RTX 6000 GPU.

\begin{figure}[t]
  \centering
  \includegraphics[width=1\textwidth]{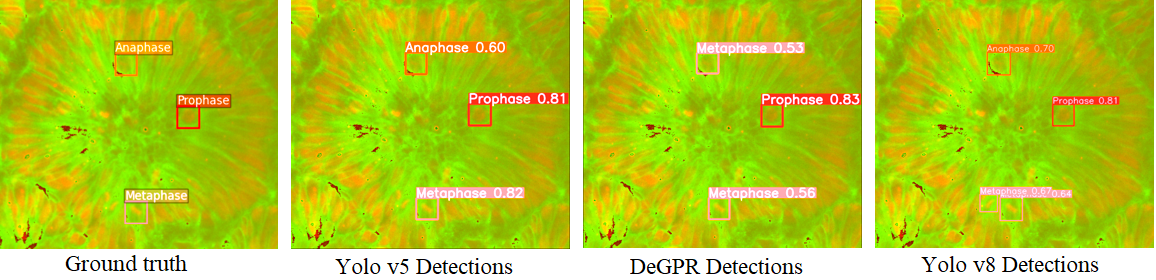} 
  \caption{Qualitative comparison of detection and cell counting models adapted for BOrg dataset. }
  \label{fig:qualitative_results2}
\end{figure}

\noindent\textbf{Results. }
Table~\ref{tab:detection_results} presents the detection and counting performance of various state-of-the-art models adapted for the BOrg dataset. Among the different model families, we observe that the efficient YOLO models significantly outperform single-stage (RetinaNet5~\cite{retinanet}) and two-stage (Faster RCNN~\cite{rcnn}) object detectors in both detection and counting tasks. The YOLO models strike a favorable balance between accuracy and efficiency, making it suitable for practical analysis tools on consumer hardware. Furthermore, despite slightly lower detection scores, DeGPR~\cite{degpr} performs well in counting tasks due to its specialized training for cell classification. Figure~\ref{fig:abalations}(a) illustrates the class-wise mean Average Precision (mAP) of the top three models, highlighting the impact of class distribution on overall performance.

To improve the models' effectiveness, we investigate the role of different preprocessing techniques and training strategies through ablation studies. Figure~\ref{fig:abalations}(b) compares the effect of various 3D-to-2D projection methods on the final mAP. We select the color mean projection based on these results for subsequent analysis. Additionally, Figure~\ref{fig:abalations}(c) evaluates three different pre-training strategies for DeGPR, with the combined pre-training on both BOrg and MoNuSAC datasets yielding the best performance.

Overall, our results demonstrate the effectiveness of adapting off-the-shelf object detection and cell counting methods to the BOrg dataset, showcasing their potential to aid researchers in understanding embryonic brain development and associated disorders through automated analysis of mitotic events.

\section{Conclusion}

We introduce BOrg, a novel brain organoid-based dataset designed to support the analysis of early brain development. This dataset comprises images collected via confocal microscopy of on-chip brain organoids that recapitulate critical aspects of the embryonic development of the human brain in its early stage. Subsequently, the data is annotated using our efficient annotation mechanism, which effectively employs a blend of expert and non-expert input. Finally, we adapt several object detection and cell counting models on this dataset to show its potential in assisting analysis in deciphering brain mechanisms. 

\bibliographystyle{splncs04}
\bibliography{refs}

\end{document}